%% file: 8FSM_final3_arxiv.tex
\documentclass[a4paper]{jpconf}
 \usepackage{cite}
\usepackage{array}
\usepackage{amsmath}
\usepackage{graphicx}
\usepackage{latexsym}
\usepackage{amsfonts}
\usepackage{pifont}
\usepackage{stmaryrd}
\usepackage{amssymb}
\usepackage{array}
\usepackage{epsfig}
\usepackage[labelformat=default, labelsep=default, justification=centerlast,  font=footnotesize]{caption}
\usepackage{verbatim}
\usepackage{multirow}
\usepackage{rotating}
\usepackage{amsthm}
\usepackage{enumerate}
\usepackage{color}
\usepackage[colorlinks]{hyperref}
\usepackage{geometry}
\usepackage{nopageno}

\newcommand\id{\leavevmode\hbox{\small1\kern-3.3pt\normalsize1}}

\def\f12{\frac{1}{2}}

\def\ms{m$\cdot$s }

\newtheorem*{definition*}{Definition}

\definecolor{darkpastelgreen}{rgb}{0.01, 0.75, 0.24}

\input{thesis_defs}

\begin{document}

\title{General relativistic effects in quantum interference of ``clocks''}

\author{M Zych$^1$, I Pikovski$^{2,3}$, F Costa$^1$ and \v{C} Brukner$^{4, 5}$}
\address{$^1$ Centre for Engineered Quantum Systems, School of Mathematics and Physics, The University of Queensland, St Lucia, QLD 4072, Australia }
\address{$^2$ ITAMP, Harvard-Smithsonian Center for Astrophysics, Cambridge, MA 02138, USA}
\address{$^2$ Department of Physics, Harvard University, Cambridge, MA 02138, USA}
\address{$^4$ Institute for Quantum Optics and Quantum Information, Austrian Academy of Sciences, Boltzmanngasse 3, A-1090 Vienna, Austria.}
\address{$^5$ Faculty of Physics, University of Vienna, Boltzmanngasse 5, A-1090 Vienna, Austria}
\ead{m.zych@uq.edu.au}

\begin{abstract}
Quantum mechanics and general relativity have been each successfully tested in numerous experiments. However, the regime where both theories are jointly required to explain physical phenomena remains untested by laboratory experiments, and is also not fully understood by theory. This contribution reviews recent ideas for a new type of experiments: quantum interference of ``clocks'', which  aim to  test novel quantum effects that arise from time dilation.  ``Clock'' interference experiments could be realised with atoms or photons in near future laboratory experiments.
\end{abstract}

\section{Introduction}
\label{sec:intro}

Quantum mechanics and general relativity have successfully passed numerous experimental verifications. Yet, all experiments that detected gravitational effects on quantum systems are still fully consistent with the non-relativistic, Newtonian theory \cite{cow, nesvizhevsky:2002quantum, jenke:2011}. Similarly, in the present-day tests of general relativity the degrees of freedom displaying general relativistic effects can consistently be described by classical physics \cite{Will:2014, PoundRebka:1960, HafeleKeating:1972a, HafeleKeating:1972b, Wineland:2010, Ye2015}. Therefore, there is a fundamental interest in finding feasible experiments that could probe the regime where the relevant degrees of  freedom requite both quantum and general relativistic description.

This contribution reviews recent proposals for testing the overlap between quantum theory and general relativity with composite quantum particles, which have dynamical internal degrees of freedom (DOF). Crucially, such systems 
can be seen as ideal ``clocks''.  Their study has revealed novel quantum effects from the relativistic time dilation arising  in interference experiments with such quantum ``clocks''  \cite{Zych:2011, Zych:2012, pikovskiuniversal2015, ZychBrukner2015arXiv, Brodutch:2015.Photons,  GoodingUnruh:2014.Self, gooding2015bootstrapping}.
According to  quantum theory any system can propagate along multiple paths in superposition, while general relativity predicts that the time elapsed for a system depends on its path, which is simply the effect of time dilation. Therefore, in the regime where both theories apply, time dilation will generally entangle the internal states of a ``clock'' to the path.  In interference experiments with ``clocks'' this will cause loss and revivals (for periodic ``clocks'') of the interference fringes \cite{Zych:2011}. Such experiments could be realised with atoms or molecules, with their electronic or vibrational energy levels as ``clocks'',  neutrons or electrons -- with ``clock'' implemented in spin precession, etc. Analogous effects arise also for photons \cite{Zych:2012} due to the Shapiro effect \cite{ref:Shapiro1964}.
Moreover, the interferometric visibility is  affected  stronger for larger systems. For macroscopic systems the loss of visibility can become very rapid and, for all practical purposes, irreversible -- time dilation can thus result in decoherence \cite{pikovskiuniversal2015}. 

The prime motivation to study the effects of proper time on quantum systems is the fact that they are particularly promising for laboratory experiments: Already current techniques allow for interference experiments with molecules comprising more than 800 atoms \cite{ref:Eibenberger2013}.
More generally, understanding how general relativity can affect quantum experiments will simply be crucial for many future tests aimed at preparing and detecting quantum states of macroscopic systems  \cite{Schmole:2016mde, Schnabel2015, Kovachy2015}. Future large-scale quantum technologies involving e.g.\ quantum communication through satellites \cite{Scheidl:2012un, Scheidl:2013dx} will also have to account for general relativistic effects -- as much as the present-day classical technologies, such as the Global Positioning System, have to incorporate such effects for their correct functioning.  The impact of general relativity on future quantum communication and information technologies is already today an active research field \cite{MannRalph:2012}. Finally, tests of general relativity are performed with ever increasing precision \cite{Ye2015} and quantum effects will unavoidably become relevant in future tests at some scale.

\section{Composite particles as ``clocks''}
\label{sec:comp_clocks}
For a relativistic particle the square of its four momentum $p^\mu$, $\mu=0,..,3$ is an invariant quantity, which describes the energy of the particle in its rest frame \cite{WeinbergGR}: $-H_{rest}c^2=\sum p^{\mu}g_{\mu\nu}p^{\nu}$, where $g_{\mu\nu}$ is the metric tensor, with signature $(-, +,+,+)$, and  $c$ is the speed of light. For a composite system, $H_{rest}$ comprises not only the rest mass $mc^2$ but also any binding or kinetic energies of the internal degrees of freedom --  the system's entire internal Hamiltonian $H_{int}$ -- and thus $H_{rest}=mc^2+H_{int}$. The energy in an arbitrary reference frame is given by the $p_0$ component and for a static symmetric metric reads $H\equiv cp_0= \sqrt{-\go(c^2\pj{}\Pj{} +\r^2)}$.

Legendre transform of $H$ yields the Lagrangian $L=L_{rest}\sqrt{-\gmn\dXmu{}\dXnu{}}/c$, where $L_{rest}$ is the rest frame Lagrangian of the composite system (and a Legendre transform of $H_{rest}$), $x^{\mu}(t)$ are the coordinates of the system's world line, and the over-dots denote  derivative with respect  to the coordinate time $t$. Furthermore, the proper time element along $x^{\mu}(t)$ reads $d\tau := \sqrt{-\gmn d\Xmu{}d\Xnu{}}/c$ and  $\sqrt{-\gmn \dXmu{}\dXnu{}}/c\equiv\td$ is simply the ``speed'' of proper time with respect to the coordinate time $t$.  This yields $L\equiv L_{rest}\td$. The action of a relativistic composite system along a world line $\gamma$ is therefore $S=\int_{\gamma} L_{rest}\td dt\equiv \int_{\gamma} L_{rest} d\tau$, whereas the proper time elapsing along $\gamma$ is given by $\int_{\gamma}\td dt\equiv \int_{\gamma} d\tau$. Thus, internal degrees of freedom of a composite point-like system evolve according to  proper time along the system's world line -- ``measuring'' the elapsing proper time. In this sense relativistic point-like  particles with internal dynamics can be seen as ideal clocks.

Any physical system thus far used as a clock in tests of time dilation was of course not fundamentally point-like. For the above framework to apply it is only required that the additional effects on the system due to its finite size are negligible. This is the case if the linear dimension of the system is sufficiently small compared to the length scale where the space-time curvature is non-negligible. In general, the higher the precision of the clock, the smaller the four-volume it can occupy before it becomes necessary to include its finite size, see also ref.~\cite{ZychPhD2015}. The above conditions also allow to define a generally covariant notion of the centre of mass for an extended system \cite{Dixon:1964}.

To the lowest order in the post-Newtonian expansion (for a static symmetric metric) \cite{poisson2014gravity} the Hamiltonian for a composite system reads \cite{pikovskiuniversal2015}
\begin{equation}
H = H_{\text{ext}} + H_{int}\left(1 + \frac{\Phi(x)}{c^2}  - \frac{p^2}{2m^2 c^2}\right),
\label{Hamiltonian}
\end{equation}
where $H_{\text{ext}}=mc^2 + p^2/2m + m \Phi(x)+f(p, \Phi)/c^2$ describes the external DOFs and $f(p, \Phi)/c^2$ incorporates the relevant relativistic corrections and any relevant external forces (e.g.~required for trapping the ``clock'').
In the above approximation
\be
\td\approx1 + \frac{\Phi(x)}{c^2}  - \frac{v^2}{2c^2}
\ee{td}
and the interaction terms $-H_{int}\frac{p^2}{2m^2c^2}$ and $ H_{int}\frac{\Phi}{c^2}$ in eq.~\eqref{Hamiltonian} can immediately be recognised as special relativistic (velocity dependent) and gravitational time dilation of internal dynamics, respectively.

The classical theory of composite particles can be canonically quantised to yield a Hamiltonian description of a quantum ``clock'' in curved space-time (or, equivalently, a path integral quantisation of the clock action can be carried out). Note, that in quantum theory both external and internal DOFs are described by operators. Since the particle in general propagates along multiple world lines in superposition, the interactions describing time dilation in quantum theory lead to entanglement between internal and external degrees of freedom of the ``clock''.  This yields the novel phenomena in interference experiments with composite systems \cite{Zych:2011, Zych:2012, pikovskiuniversal2015} which we review in this contribution.

The two terms describing time dilation in eq.~\eqref{Hamiltonian} can also be justified  by considering the mass-energy equivalence (for the mass of the particle and its internal energy) \cite{Einstein:1905, Einstein:1911}. It entails that changing the particle's internal energy by $E$ changes its mass $m$ by $E/c^2$, i.e.\  $m\rightarrow m+E/c^2$. 
The principle applies to any internal energy eigenstate and in quantum theory, due to its linear structure, the principle also applies to any superposition of internal energy eigenstates, thus in quantum theory $m\rightarrow m\id+H_{int}/c^2$,  where $\id$ is the identity operator on the state space of the internal degrees of freedom.  Incorporating the above into the non-relativistic quantum Hamiltonian $mc^2+\frac{p^2}{2m}+m\phi$ yields the interaction terms present in \eqref{Hamiltonian}, see ref.~\cite{ZychBrukner2015arXiv} {for further details and for the new possible tests of the Einstein Equivalence Principle in quantum theory.


\section{Interference of ``clocks''}
\label{sec:clock_interference}
So far time dilation was tested by comparing time measured by clocks that followed different, effectively classical, world lines \cite{HafeleKeating:1972a, HafeleKeating:1972b, Wineland:2010, Ye2015}. Such tests can be seen as realisations of the famous ``twin paradox'', a thought experiment where twin siblings embark on separate voyages from the same event in space-time and  when their world lines meet again, they discover that they have aged differently. The age of each twin depends on the proper time elapsed along his/her path. Here we discuss a quantum version of the time dilation experiments: a single ``clock'' (a particle with internal dynamics, e.g.\ an atom in a superposition of two  internal energy levels) which follows in superposition two world lines along which different proper time elapses \cite{Zych:2011}. This can  be seen as a quantum version of the twin paradox, where a ``quantum only child''  embarks on two different voyages in superposition and when the world lines are recombined, the traveller  has aged in superposition by different amounts of time -- becoming older-and-younger than himself, in a superposition.


For the realisation of the ``quantum twin paradox'' consider a Mach-Zehnder interferometer where the quantum ``clock''  follows in superposition two fixed paths\footnote{We neglect the proper time effects due to the finite spread of the ``clock's'' wave function along each path as compared to the effects stemming from the  proper time difference between the two paths. This is justified for the experiment realised with massive systems like atoms or molecules.} $\gamma_1$ and $\gamma_2$, see figure~\ref{machzehnder}. %
\begin{figure}[h!]
 \begin{minipage}[l]{0.5\textwidth}
\hspace{-1cm}
\includegraphics[width=8.0cm]{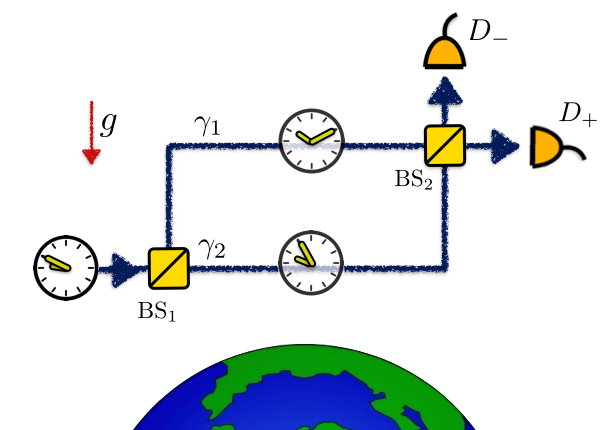}
\end{minipage}\hfill
  \begin{minipage}[r]{0.5\textwidth}
\caption{\footnotesize Mach-Zehnder interferometer for testing the quantum version of the twin paradox with interfering ``clocks''. The setup consists of two beam splitters (BS$_{1,2}$), two detectors $D_\pm$ and is stationary in a laboratory frame on earth, in a gravitational field $g$. The interferometric paths $\gamma_1$ and $\gamma_2$ are arranged such that different proper time elapses along each of them. The ``clock'' can be any small particle with time-evolving  internal degrees of freedom.  Such an interference experiment will not only display a phase shift, but also (periodic -- for a periodic ``clock'') modulations of the interferometric visibility: The interferometric visibility will be diminished to the extent to which the information about the path becomes available from the ``clock'' due to time dilation.\hspace*{\fill}}
\label{machzehnder}
\end{minipage}
\end{figure}
%
%
The state of the particle inside the setup can be described by a superposition $\frac{1}{\sqrt{2}} \left(i|\Psi_{1}\ket +  |\Psi_{2}\ket\right)$
 where the states $|\Psi_1\ket$, $|\Psi_2\ket$ are associated with the paths $\gamma_1$, $\gamma_2$, respectively. 
 With the Hamiltonian \eqref{Hamiltonian}  we can write these states explicitly
\begin{equation}
\label{psi_i}
|\Psi_{i}\ket=e^{-\frac{i}{\hbar}\int_{\gamma_i}dt\; \left[H_{ext}+H_{int}\left( 1- \frac{p^2}{2m^2c^2}+\frac{\Phi(x)}{c^2}\right)\right]}|x^{in}\ket|\tau^{in}\ket.
\end{equation}
where $|x^{in}\ket$ and $|\tau^{in}\ket$ denote the initial states of the external and the internal DOFs, respectively. 
Using eq.~\eqref{td} the internal state evolving along $\gamma_i$ can be written as  
\be
|\tau_i\ket=e^{-\frac{i}{\hbar}\int_{\gamma_i}dt\; H_{int}\td}|\tau^{in}\ket,
\ee{tau_i}
where we used $\frac{v^2}{2c^2}\equiv \frac{p^2}{2mc^2}$ (for a particle on a fixed path, in the small velocity regime). This can be further simplified into $|\tau_i\ket=e^{-\frac{i}{\hbar}\int_{\gamma_i}d\tau\; H_{int}}|\tau^{in}\ket$ where we explicitly see that the internal state of the particle following path $\gamma_i$ evolves according to the proper time along that path. 
In the interferometric experiment the paths are finally recombined,  which yields the following probabilities for registering the particle in the detectors $D_\pm$
\be
P_{\pm}=\frac{1}{2}\pm\frac{1}{2}|\langle \tau_1|\tau_2\rangle|\cos(\Delta\phi),
\ee{probab}
where $\Delta\phi$ is the total relative phase, which can also be explicitly evaluated from eq.~\eqref{psi_i}, see also sec.~\ref{sec:discussion}.
In  eq.~\eqref{probab}, the quantity $\mathcal V=|\langle \tau_1|\tau_2\rangle|$ describes the visibility of this interference pattern and explicitly reads
\be
\mathcal V=|\bra e^{-\frac{i}{\hbar}H_{int}  \Delta \tau } \ket|,
\ee{visib}
where $\Delta \tau:=\int_{\gamma_2}d\tau - \int_{\gamma_1}d\tau$ is the proper time difference between $\gamma_1$ and  $\gamma_2$,  
and the expectation value is taken with respect to the initial internal state of the ``clock'', here $|\tau^{in}\ket$.

The visibility $\mathcal{V}$ remains maximal if there is no time dilation between the interfering paths, in particular it is always maximal in the non-relativistic limit (when eq.~\eqref{Hamiltonian} reduces to $H_{int}+ p^2/2m + m \Phi(x)$). On the other hand,  $\mathcal{V}$ vanishes in the classical theory -- even in curved space-time --  no interference takes place for classical particles.  
Modulations in the visibility described by eq.~\eqref{visib}, are a quantum effect that arises only due to the relativistic time dilation and can thus test the regime where both quantum and general relativistic phenomena play a role.

Equation~\eqref{visib} quantifies to which extent an internal state evolving along $\gamma_1$ becomes distinguishable from that evolving along $\gamma_2$ due to time dilation.  If the ``clock'' has a characteristic evolution time $t_\perp$ after which its initial state evolves into an orthogonal one,  also called the orthogonalisation time, the visibility vanishes if $\Delta\tau=t_{\perp}$. Both $\mathcal V$ and $t_{\perp}$ can be expressed in terms of internal energy moments $ \bra H_{int}^n\ket$:
\be
 \mathcal{V}=\left|\sum_{n=0}^\infty  \left(\frac{-i\Delta \tau}{\hbar}\right)^n \bra H_{int}^n\ket\right|
 \ee{visib_moments}
 and $\pi\hbar/t_\perp \leq (2\bra H_{int}^n \ket )^\frac{1}{n}$, for $n >0$~\cite{zielinski}, which is valid also for mixed states.
Keeping only up to second moments in the right-hand side of eq.~\eqref{visib_moments} gives:
\be
\mathcal{V}\approx\sqrt{1- \left(\Delta \tau  \Delta H_{int}/\hbar\right)^2},
\ee{visib_variance}
where $\Delta H_{int}=\sqrt{\bra H_{int}^2\ket-\bra H_{int}\ket^2}$ is the internal energy variance. In the same approximation  $t_\perp \geq \frac{\pi \hbar}{2\Delta H_{int}}$  \cite{mandelstam, fleming, margolus, kosinski}.

The which-way  information revealed by the ``clock'' can be quantified by the distinguishability of the paths, here: $\mathcal D = \sqrt{1-|\bra\tau_1|\tau_2\ket|^2}$, see \cite{Zych:2011} for details. This yields  the well-known quantitative expression of quantum complementarity: $\mathcal V^2+\mathcal D^2=1$, see~\cite{wooters, green_yasin, englert}.

Equation~\eqref{visib} has been derived here for pure states but it directly generalises to mixed internal states\footnote{The most general initial state to which eq.~\eqref{visib} applies is $\rho(0)=\rho_{cm}(0)\otimes \rho_{int}(0)$ with $\rho_{cm}(0)=|\psi_{cm}(0)\ket\bra \psi_{cm}(0)|$, where $|\psi_{cm}(0)\ket$ is a superposition of two path modes just after BS$_1$.} \cite{pikovskiuniversal2015} -- it only depends on the internal energy distribution, not on the purity of the state.
For pure internal states the internal energy distribution gives the rate of internal state change (given by $t_\perp$) and the visibility modulations can be interpreted as due to which-way information that becomes available from  the ``clock''. Mixed states, on the other hand, in general cannot be directly interpreted as ``clocks'' -- e.g.~thermal states are stationary -- and visibility modulations have two equivalent interpretations: as time dilation of internal dynamics or redshift of internal energy values, 
 see ref~\cite{pikovskiuniversal2015} for further discussion. For mixed states in general one has loss of visibility without the gain in which-way information, which might become classically ``scrambled'', and $\mathcal V^2+\mathcal D^2\leq1$.

Finally, we stress that in the relativistic scenario considered here, the visibility can remain maximal only if the internal state of the interfering  particle is an exact eigenstate of its internal Hamiltonian. Such a state only evolves by a global phase and can be seen as a ``switched-off clock''.

\section{Interferometric visibility -- examples}
Here we discuss interferometric visibility in three experimentally promising implementations of the ``clock'',  and where additionally eq.~\eqref{visib} admits a closed analytic expression.

\textbf{\textit{Periodic ``clocks''}} Consider a ``clock''  periodically evolving between just two mutually orthogonal states with a period $2t_\perp$. We expect $\mathcal V=0$ for $\Delta\tau=(2k+1)\,t_\perp$, $k=0,1,2,...$, when the ``clock'' stores maximal which-way information, and  $\mathcal V=1$ for $\Delta\tau=2k\,t_\perp$ -- when no which-way information is available.  In general for a periodic ``clock'' eq.~\eqref{visib} reads \cite{Zych:2011}
\be
\mathcal V=\left|\cos\left( \frac{\pi \Delta\tau}{2 t_\perp} \right)\right|.
\ee{visib_2level}
Such a periodic ``clock'' can be realised as a superposition of two energy eigenstates $\frac{1}{\sqrt{2}}(|E_1\ket+|E_2\ket)$ for which $t_\perp=\pi \hbar/|E_1-E_2|$. Note that $\nu=2\pi/2t_\perp\equiv$ is simply the angular frequency of this ``clock''.
A  physical implementation of such a periodic ``clock'' is  possible e.g.~in internal energy levels of an atom, as is routinely employed in atomic clocks.

\textbf{\textit{Continuous-variables: Photons as ``clocks''}}
A ``clock'' can also be implemented in the position of a photon. For a fixed  observer, light passing closer to a massive object moves slower than light passing further from the mass -- this effect is known as Shapiro delay \cite{ref:Shapiro1964} (and has been tested in the classical regime by measuring round-trip times of radar \cite{ref:Shapiro1964, ref:Shapiro1971} and radio waves \cite{Bertotti:2003rm} in the Solar System). If a single  photon is used in the setup in figure\ref{machzehnder}, the amplitude following path $\gamma_2$ is predicted to reach the final beam splitter ($BS_2$) later than then the amplitude which followed $\gamma_1$, even for paths of equal local lengths\footnote{The path lengths can be  calibrated by maximising interference contrast when the interferometer is placed  such that both paths are at the same gravitational potential.}. For a Gaussian distribution of frequencies in the  photon wave-packet $f(\nu)=(\frac{a^2}{\pi})^{1/4}\exp({\frac{-a^2}{\;2}(\nu-\nu_0)^2})$ the visibility reads \cite{Zych:2012} 
%
$\mathcal{V}=e^{-\left(\frac{\Delta \tau}{2a}\right)^2}.$
%
Defining as distinguishable the photon wave packets with overlap $1/e$ or smaller, the precision of a Gaussian ``clock'' is $t_\perp =2a$. Such a photon-clock is non-periodic and therefore the visibility is also a non-periodic function of the clock's precision and of the time dilation between the paths.

\textbf{\textit{Thermal states under time dilation --  mixed ``clocks''}}


Here we consider a particle in a thermal internal state: $\rho_{in}=e^{-\beta H_{int}}/\mathcal{Z}(\beta)$, where  $\mathcal{Z}(\beta)=\Tr\{e^{-\beta H_{int}}\}$ is the partition function, $\beta =1/k_BT$, $k_B$ is the Boltzmann constant, and $T$ denotes temperature, and eq.~\eqref{visib} reads 
$\mathcal{V}=\left|  \frac{\mathcal{Z}(\beta +{i\Delta \tau}/{\hbar})}{\mathcal{Z}(\beta)} \right|$ (note that $\beta+i\Delta \tau/\hbar$ is effectively a ``complex temperature'').
As an example we consider that internal DOFs comprise $N$ independent harmonic modes (such as normal modes of a molecule, a nanosphere, or a bulk macroscopic system) \cite{pikovskiuniversal2015}. In this case $H_{int}=\sum_{i=1}^N (n_i+\f12)\hbar \omega_i$, where  $n_i$ are the number operators, and $\omega_i$ -- the mode frequencies and the visibility reads
\be
\mathcal{V}=\Pi_{i=1}^N \left|  \frac{1-e^{-\beta \hbar\omega_i}}{1-e^{-\left(\beta +\frac{i\Delta\tau}{\hbar}\right)\hbar\omega_i}} \right|,
\ee{visib_harmonic}
In a large temperature limit, $1/\beta>>\hbar\omega_i$, the energy variance of a harmonic mode is $\beta^{-1}$ and in the approximation~\eqref{visib_variance} the visibility 
reads  $\mathcal{V}\approx\left(1- ({\Delta\tau}/{\beta\hbar})^2\right)^{N/2}$, which for large $N$ becomes
\be
\mathcal{V}(t) \approx e^{-\left(\sqrt{\frac{N}{2}}\frac{k_BT\Delta\tau}{\hbar }\right)^2}.
\ee{visib_approx}
As much as for pure internal states the visibility can be expressed in terms of their orthogonalisation time, for thermal states it can be expressed in terms of heat capacity, in the Einstein's model given by $(\Delta H_{int})^2/(k_B T^2) \approx N k_B$,
see ref.~\cite{pikovskiuniversal2015} for further details; and ref.~\cite{CarlessoBassi:2016} for the analysis using the Debye model of solids.

\section{Gravitational and inertial time dilation effects}
\label{time dilation}
Time dilation between the paths of an interferometer will in general depend both on inertial and gravitational effects, $\Delta\tau=\oint dt \left(1- \frac{p^2}{2m^2c^2}+\frac{\Phi(x)}{c^2}\right)$, where $\oint\equiv \int_{\gamma_1}- \int_{\gamma_2}$, in full analogy to classical tests \cite{HafeleKeating:1972a, HafeleKeating:1972b}. Here we discuss two different geometries of the paths that allow isolating either the effects of gravitational time dilation or the special relativistic (inertial) time dilation.

\textbf{\textit{Gravitational time dilation -- lifting the ``clock''}}
 Consider the paths as sketched in figure~\ref{machzehnder}, where acceleration and deceleration in the vertical direction, as well as horizontal velocity (which might also be zero when the system is trapped) are the same for both $\gamma_1$ and $\gamma_2$. This assures that inertial effects give the same contribution to the proper time elapsing along both trajectories, and the proper time difference stems from the gravitational potential alone. In the approximation of homogeneous gravitational field with acceleration $g$ we thus obtain
\be
\Delta\tau=\frac{ght}{c^2},
\ee{dtau_grav}
where $h$ is the height difference between the paths and $t$ is the laboratory time which the ``clock'' spends at the fixed heights.
\begin{figure}[h]
  \begin{minipage}[l]{0.45\textwidth}
  \hspace{-0.2cm}  \includegraphics[width=0.99\textwidth]{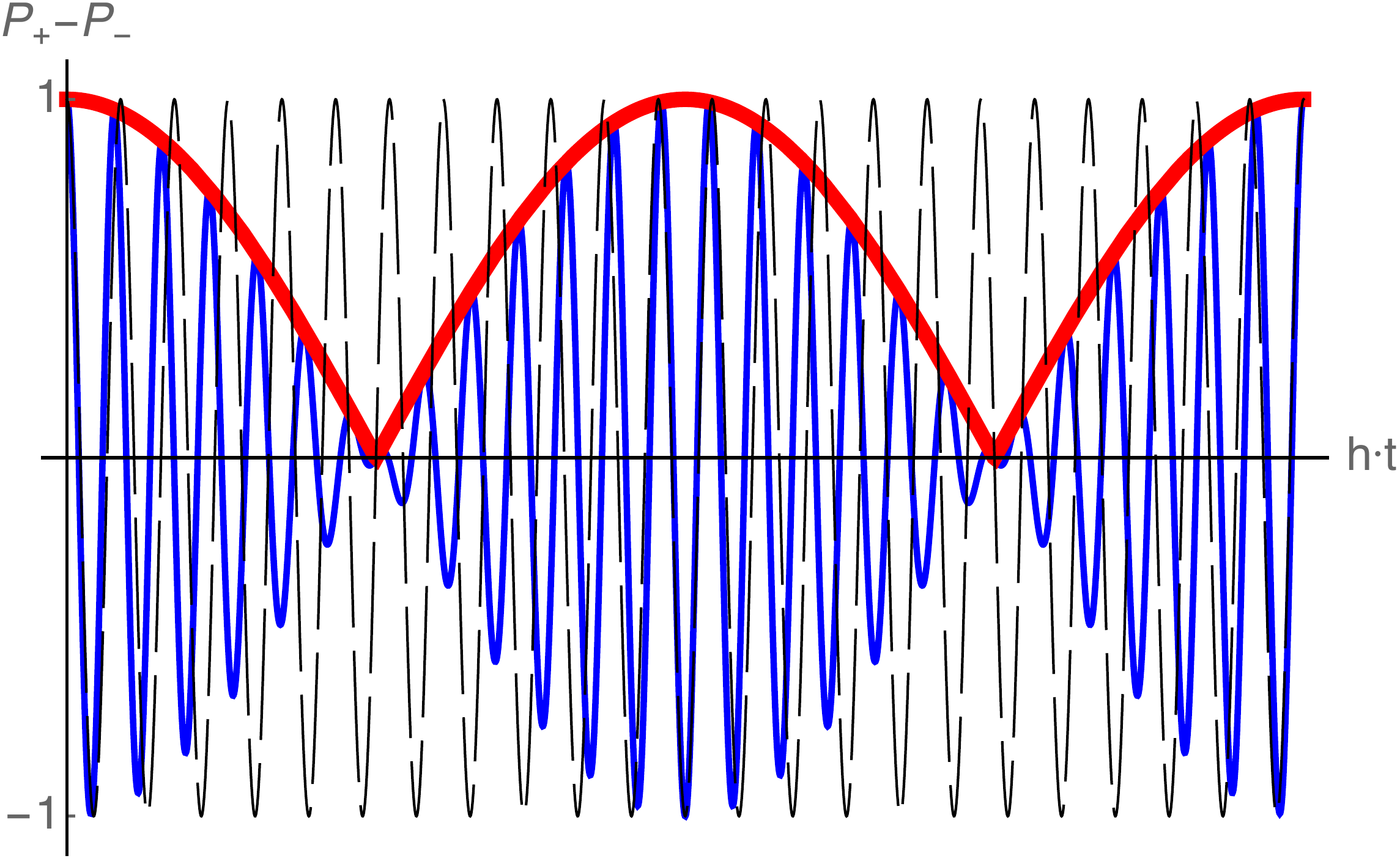}
  \end{minipage}\hfill
   \hspace{-0.2cm}\begin{minipage}[r]{0.55\textwidth}
    \caption{\footnotesize   ``Clock'' interference under gravitational time dilation, plotted as a function of the interferometer area $h t$ [m$\cdot$s], where $h$ is the height difference between the paths and $t$ is the time for which the ``clock'' is kept in superposition. In a non-relativistic theory only a relative phase is acquired but the visibility remains maximal (dashed, black line), as well as in a relativistic theory for  a ``switched-off clock'' (particle in an internal energy eigenstate).  In a relativistic case with a ``switched-on'' ``clock'' (particle in a superposition of internal energy levels) the visibility periodically drops and revives (thick red line), and a correction to the phase shift arises (blue line). Both quantum theory and the notion of gravitational time dilation are necessary to explain the visibility modulations in this experiment.\hspace*{\fill}} \label{fig:probab}
  \end{minipage}
\end{figure}
Figure~\ref{fig:probab} shows interference pattern predicted for the experiment realised with a  periodic ``clock'' under gravitational time dilation.
In an earth based experiment the only parameter defining $\Delta\tau$ in eq.~\eqref{dtau_grav} is the space-time area $ht$ enclosed by the paths of the interferometer and eq.~\eqref{visib_2level} reads $\mathcal V=\cos(\frac{\nu ght}{2c^2})$. For a ``clock'' with frequency $\nu=10^{15}$rad/s (which can be realised e.g.\ in Strontium)  the space-time area of the interferometer required to observe full loss and revival of the interference contrast is $\sim10\mathrm{\;m\!\cdot\!s}$, where we used $g\approx10$m s$^{-2}$ and $c\approx3\cdot10^8$m s$^{-1}$. The simple cosine law describing loss and revivals of the interference contrast is a key desirable feature of an experiment with a two-level ``clock'', as the revival of the visibility would allow for an unambiguous distinction between time dilation effects and unavoidable experimental imperfections. While such an experiment is still very challenging,  it appears within reach of next generation matter-wave interferometers \cite{Kovachy2015}.  An analog version of this test has been recently realised in interference of  Rubidium, where the ``clock'' was implemented via spin precession in an external magnetic field and  where time dilation was simulated by preparing magnetic fields of different strengths along the two paths \cite{margalit2015self}.
%
%

The visibility in such a gravitational test, but with the interfering particle in a mixed internal state, is plotted in figure~\ref{fig:visib}.
\begin{figure}[h!]
 \hspace{-0.3cm} \begin{minipage}[l]{0.5\textwidth}
   \includegraphics[width=\textwidth]{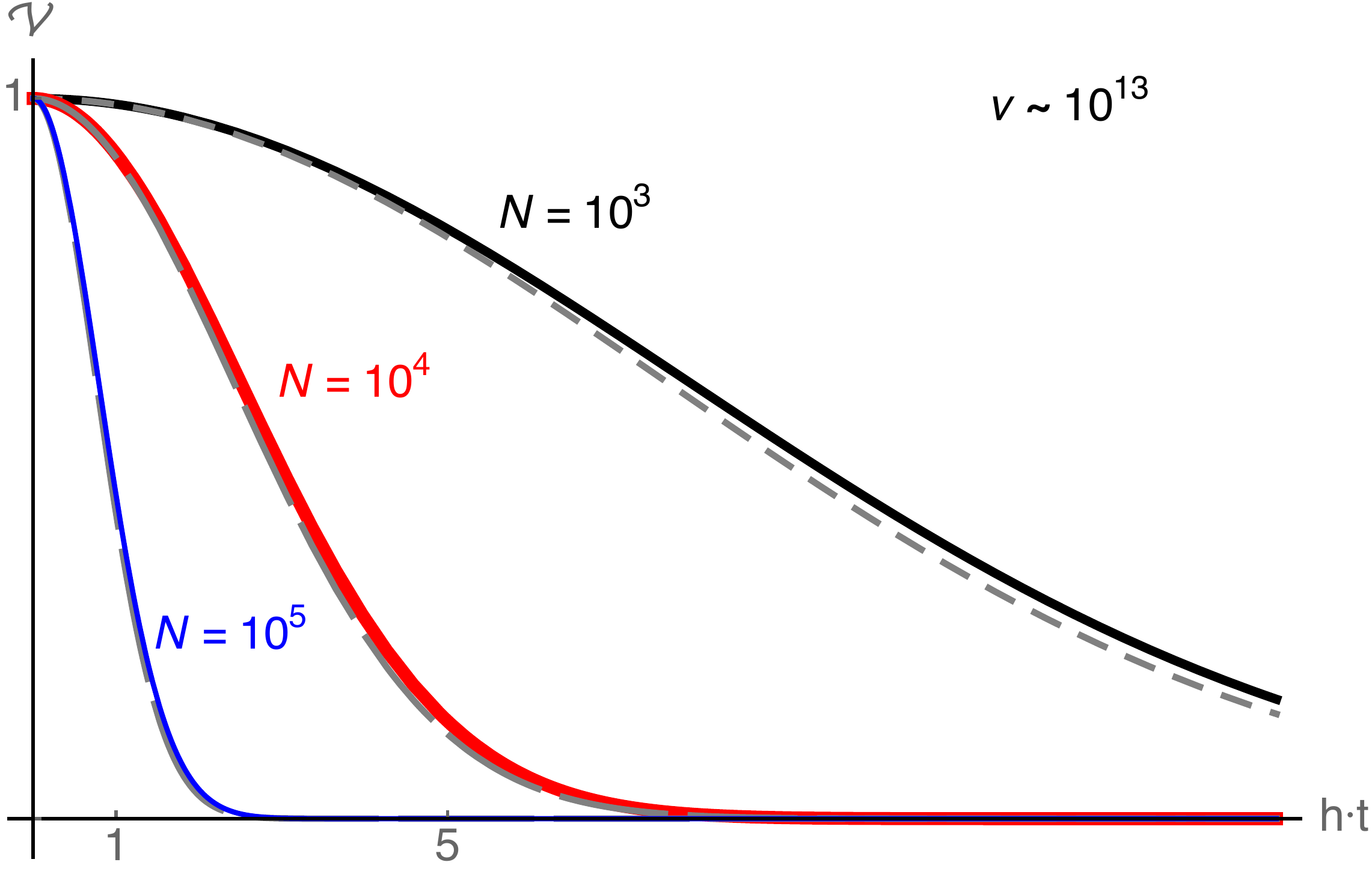}
  \end{minipage}\hfill
\begin{minipage}[r]{0.5\textwidth}
 \includegraphics[width=\textwidth]{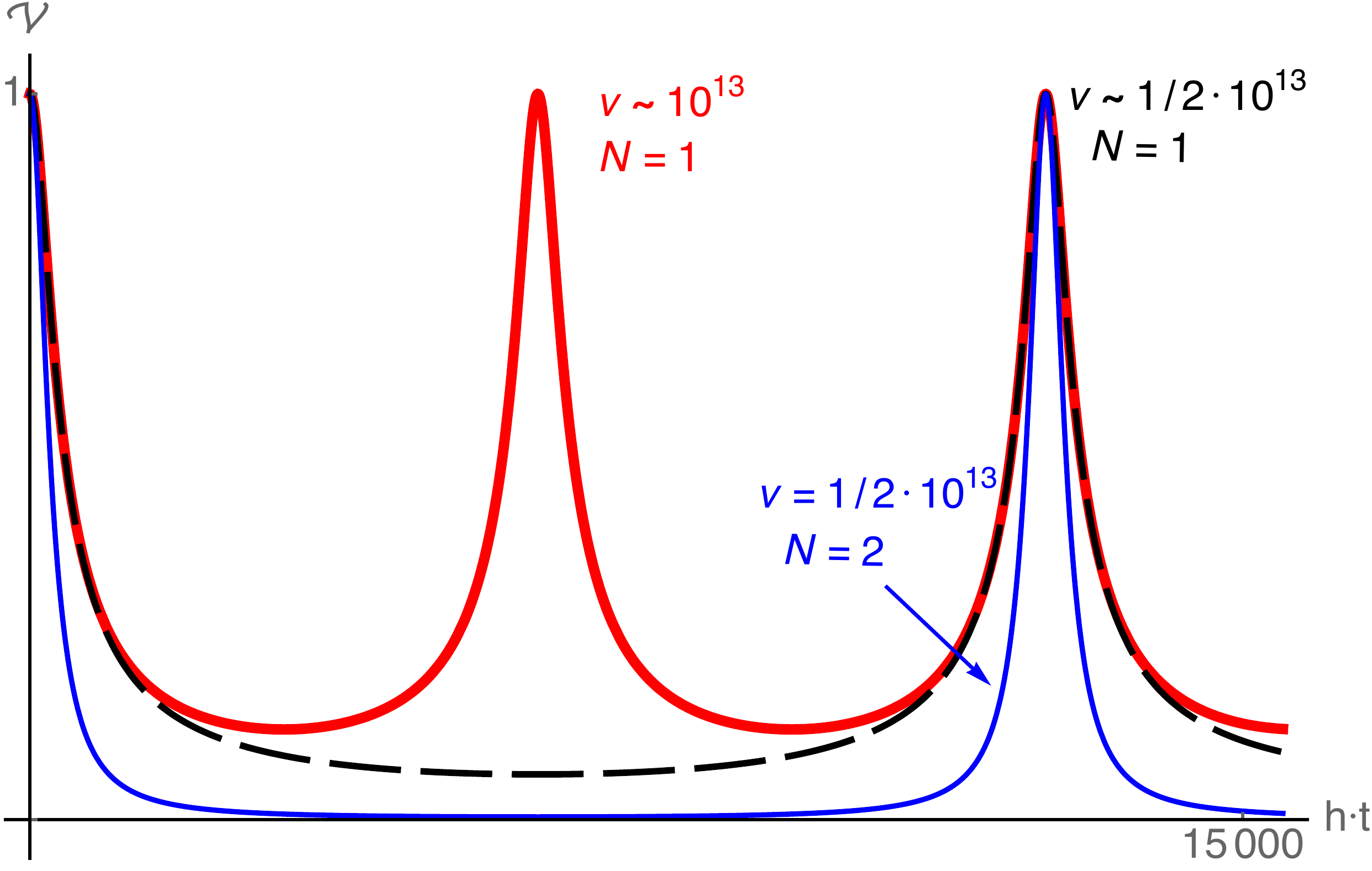}
  \end{minipage}
   \begin{minipage}[l]{1\textwidth}
\caption{\footnotesize Visibility $\mathcal V$ in an interference experiment with a particle in a thermal internal state at $T=300$K under gravitational time dilation. $\mathcal V$ is plotted as a function of the space-time area $h\!\cdot\!t$ [m$\!\cdot\!$s], with  $h$ the height difference between the paths and $t$ -- time for which the amplitudes are kept at different heights. Left: the visibility loss is faster for systems with higher number $N$ of thermalised modes. Continuous lines represent the analytic result, eq.~\eqref{visib_harmonic} for  $\nu\sim10^{13}$rad/s. Dashed lines represent the approximation obtained for high $T$ and large $N$, eq.~\eqref{visib_approx}. Right: revivals of the visibility occur for any system of finite size. The revival time is a decreasing function of the lowest frequency of the system: the thick red line shows the visibility for a single harmonic mode at frequency $10^{13}$rad/s, dashed black line -- for half that value and the thin blue line illustrates how the pattern looks for the latter frequency but two modes. Note that the revivals are shorter the more modes the system comprises.\hspace*{\fill}} \label{fig:visib}
\end{minipage}
\end{figure}
%
%
The model employed here describes the internal state as $N$ harmonic modes with equal frequency at a temperature $T=300$K. The visibility thus given by eq.~\eqref{visib_harmonic} with $\omega_i\equiv\nu$, for $i=1,..,N$ and $\Delta\tau$ as in eq.~\eqref{dtau_grav}. Revivals of the visibility occur for such a mixed state as well and  depend on the lowest frequency for the considered example (right panel of figure~\ref{fig:visib}). Moreover, the more modes the system comprises, the faster the visibility becomes negligible with increasing the superposition size $h$ or time $t$ (left panel of figure~\ref{fig:visib}); and the shorter is the duration of the revival peaks (right panel of figure~\ref{fig:visib}).
For a system with Avogadro number of particles in a superposition of $h=1$ mm, the visibility is estimated to drop below $1$\% already after $\sim 2\, \mu$s. Moreover, the frequency of the lowest phonon mode decreases with the size of the system  and therefore the revival time becomes increasingly long: For the model above with $\nu\sim500$rad/s the revival time becomes $10^{17}$s, the estimated age of the Universe. See ref.~\cite{pikovskiuniversal2015} for details on how time dilation can effectively result in decoherence. In the general case, for a system containing modes of arbitrary frequencies (as opposed to all modes at the same frequency) the revival time will be given by the fundamental frequency (the greatest common divisor of all the system frequencies), which in general will be  much longer than the period corresponding to the lowest frequency.

\textbf{\textit{Special-relativistic time dilation -- rotating the ``clock''}}
As an illustrating example, we consider a possible implementation of ``clock'' interference in the presence of special-relativistic time dilation, i.e.~due to the relative velocity between two trajectories. We analyse a scenario in which the relative velocity is due to circular motion of one of the two amplitudes.

Consider a setup where an interferometer is mounted on a rotating platform, figure~\ref{rotatingpicture}.
The ``clock'' is first sent through a beam splitter which creates a coherent superposition of two amplitudes. One amplitude is sent to a trap near the  centre of rotation, while the other is sent to a trap at a distance $R$ from the centre. The two amplitudes are kept in the respective traps for some time $t$ (as measured in the laboratory frame) and then sent back to the beam splitter, where they are recombined to produce interference fringes.
\begin{figure}[h]
  \begin{minipage}[l]{0.4\textwidth}
   \includegraphics[width=0.7\textwidth]{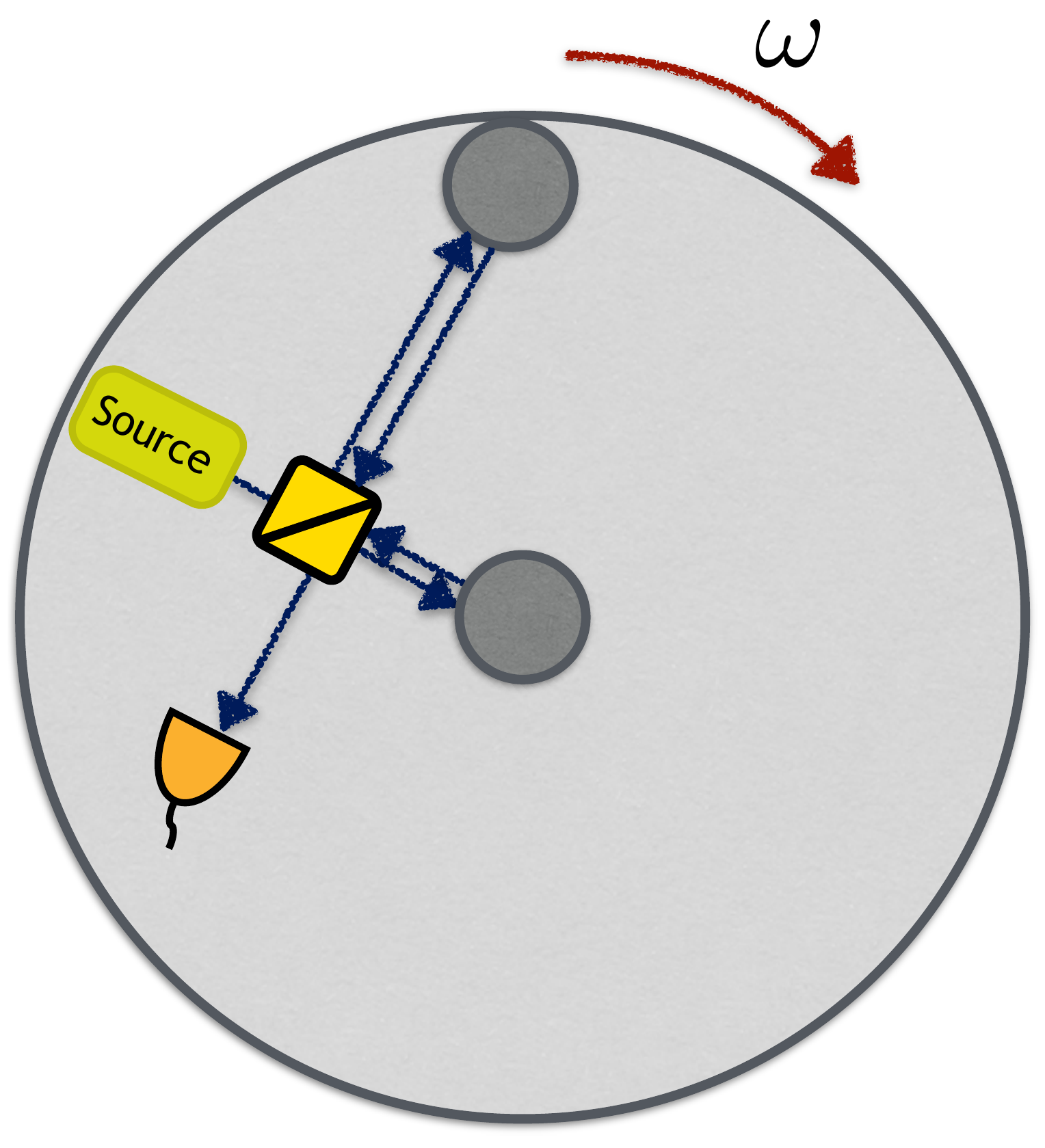}
  \end{minipage}\hfill
  \begin{minipage}[r]{0.6\textwidth}
    \caption{\footnotesize Setup for testing special-relativistic time dilation effects on interference of ``clocks''.  A particle in a superposition of internal energy levels (a ``clock'') is sent to an interferometer mounted on a rotating platform. Each amplitude is stored for a laboratory time $t$ in a trap, one near the centre and one near the rim of the platform. The two amplitudes are then recombined. The rotation velocity of the platform induces time dilation between the two traps, which affects the interferometric visibility.\hspace*{\fill}} \label{rotatingpicture}
  \end{minipage}
\end{figure}

If the platform rotates at an angular velocity $\omega$, the outer trap moves with linear velocity $v=\omega R$. Therefore, the outer amplitude accumulates a proper time $\tau\approx t(1- \frac{\omega^2 R^2}{2c^2})$, while the amplitude in the centre accumulates proper time $t$ (laboratory time), giving the proper time difference
\begin{equation}
\Delta \tau \approx t \frac{\omega^2 R^2}{2c^2}.
\label{rotatingtime}
\end{equation}
(For simplicity, we assumed that the time needed to transfer the amplitudes to the traps and back is negligible with respect to $t$.) For an equal superposition of two internal energy levels, with transition frequency $\nu=\left|E_1-E_2\right|/\hbar$, the visibility is modulated as
 \be
\mathcal V=\left|\cos\left( \frac{\nu \omega^2 R^2}{2 c^2} t \right)\right|.
\ee{rotatingvisib}

To estimate the magnitude of the effect, consider a platform rotating with angular velocity $\omega\sim100$
 rad/s, a clock with optical transition frequency $\nu\sim 10^{15}$ rad/s, and a platform with radius $R\sim 1$ m. Complete loss of visibility then occurs if the amplitudes are recombined after $t\sim 0.1$~s.

In comparison, the time dilation produced by a difference $\Delta\Phi$ in gravitational potential is $\Delta \tau \approx \Delta \Phi t/c^2$, which reproduces \eqref{rotatingtime} for the effective centripetal potential $\Delta\Phi = \frac{\omega^2 R^2}{2}$. Performing two clock-interference experiments: one in the gravitational field and the second on a rotating platform, would constitute a  new test of the equivalence principle -- in a hitherto untested regime where both external and internal DOFs of test system require relativistic as well as quantum description.

\section{Discussion}
\label{sec:discussion}
Here we address key questions discussed during the 8$^{th}$ Symposium on Frequency Standards and Metrology, (Potsdam, October, 2015).

\textbf{\textit{Why should one measure visibility and not the phase?}}
For the visibility modulation  it is crucial that internal energy contributes to the total mass as an operator. The relative phase, eq.~\eqref{probab}, can be written as\footnote{The expression is valid for any state with internal energy distribution symmetric around the mean $ \langle H_{int}\rangle$ and up to additional phases stemming from the non-gravitational forces required to control the paths,  e.g.~to trap the ``clock''.} $\Delta\phi_{GR}=(mc^2+\langle H_{int})\rangle\Delta\tau/\hbar$. Thus, first of all, a theory where only the mean internal energy  contributes to the total mass ($m\rightarrow (m +  \langle H_{int}\rangle/c^2)\id$) suffices to explain the phase shift -- but does not predict the visibility loss. Secondly,  in the non-relativistic, Newtonian, limit the phase shift does not vanish but is given by $\Delta\phi_N\approx m\Delta\Phi t/\hbar $ (for the gravitational experiment), which is fully explained by a Newtonian gravitational potential in Euclidean space-time -- with absolute time -- as confirmed in numerous experiments \cite{cow, rauch, chu}.
Moreover,  even beyond the non-relativistic limit, the phase shift $\Delta\phi_{GR}$ can be explained by a phase $\Delta \phi'_N$ stemming from a modified gravitational potential, i.e. where Newtonian gravity is modified but one still considers an Euclidian space-time.  More generally, a simple model is given in ref.~\cite{Zych:2012} that reproduces results of the existing tests of time dilation, such as~\cite{Wineland:2010, Ye2015}, as well as predicts correct relativistic phase shifts for quantum interference of ``clocks'', but which does not predict any loss of visibility.
Observation of the modulations in the visibility described by eq.~\eqref{visib}  would simultaneously disprove non-relativistic, Newtonian gravity, as well as a classical description of the world lines of the ``clocks''. This has not been achieved thus far and cannot be achieved by the measurements of the phase shift alone. 

\textbf{\textit{In atomic interferometers  there is no time dilation between the paths because atoms are in free fall.}} In the specific, Kasevich-Chu, atom interferometers
\cite{TinoKasevichSchool} proper time is indeed the same along the paths, to lowest order in relativistic corrections. This is however a specific feature of a particular setup and needs not to be true for interferometers with different geometry. For example, time dilation will not vanish in interferometers with asymmetric paths, or when using Bloch lattices \cite{TinoFermiSchool}. Importantly, the issue of vanishing proper time difference is unrelated to whether the paths are in free fall or not. Time dilation can be non-vanishing even between two free falling world lines, and interferometric visibility for this scenario has been studied in \cite{gooding2015bootstrapping}.

\textbf{\textit{In full analogy to spin echo it is always possible to ``reverse'' the effect of time dilation and ``bring the coherence back''.}} In order to have high interference contrast the paths must be arranged such that the proper time difference between them is small compared to the characteristic time scale of the system (which in general depends on its internal Hamiltonian and state). For a two-level atom such a ``gravitational echo'' could be a feasible way to assure high visibility. However, when the internal energy spread grows, e.g~ due to large number of internal modes, even small fluctuations in the interferometer area lead to a substantial loss of visibility,  figure~\ref{fig:visib} (right). For increasingly composite systems, the precision with which the paths need to be controlled thus grows, but this precision will be limited due to the Heisenberg uncertainty. For sufficiently large systems time dilation induced loss of visibility can become for all practical purposes irreversible, even though it stems from fundamentally reversible unitary dynamics.

For a discussion of time dilation in quantum theory and additional topics, we refer the reader to ref.~\cite{Pikovski:2015wwa}.

\section{Conclusion}
\label{sec:conclusion}
Quantum effects have been demonstrated with complex systems comprising hunderds of atoms \cite{arndt, ref:Hackermuller2004, gerlich, ref:Eibenberger2013, haslinger2013universal}. The regime  where general relativity affects internal dynamics of such systems might soon allow testing the interplay between quantum mechanics and general relativity.  Composite quantum systems also offer intriguing new insights into the notion of time in quantum theory: They can describe a  clock that runs different proper times in superposition and show that clocks generally must become entangled due to their unavoidable gravitational interactions \cite{Castro2015arXiv}. 
Finally, considering systems with quantised  internal energy calls for a suitable quantum formulation of the Einstein Equivalence Principle (EEP) and its new quantum tests \cite{ZychBrukner2015arXiv}.  Composite quantum particles subject to relativistic effects are particularly suited for exploring the joint foundations of quantum theory and general relativity.

\ack{
M.Z. would like to  thank the organisers of the 8$^{th}$ Symposium on Frequency
Standards and Metrology 2015 for hospitality and participants of the conference for discussions, 
This work was supported by the NSF through a grant to ITAMP, the Austrian Science Fund (FWF) through the Special Research Program Foundations and Applications of Quantum Science (FoQuS) and Individual Project (No.\ 2462), the Australian Research Council Centre of Excellence for Engineered Quantum Systems through grant number CE110001013 and by the Templeton World Charity Foundation, grant TWCF 0064/AB38.}

\end{document}

%% file: thesis_defs.tex
\newcommand{\pj}[1]{p_{#1j}}
\newcommand{\Pj}[1]{p_{#1}^j}

\newcommand{\Xmu}[1]{x_{#1}^\mu}

\newcommand{\Xnu}[1]{x_{#1}^\nu}

\newcommand{\dXmu}[1]{\dot x_{#1}^\mu}

\newcommand{\dXnu}[1]{\dot x_{#1}^\nu}

\def\gmn{g_{\mu\nu}}

\def\go{g_{00}}

\def\r{H_{rest}}
\newcommand{\bra}{\langle}
\newcommand{\ket}{\rangle}

\def\td{\dot\tau}

\newcommand{\be}{\begin{equation}}
\newcommand{\ee}[1]{\label{#1} \end{equation}}
\newcommand{\bbe}{\begin{equation*}}
\newcommand{\eee}{\end{equation*}}

\def\Tr{ {\textrm{Tr} }}

\def\f12{\frac{1}{2}}

\def\ms{m$\cdot$s }